\documentclass[aps,preprint,showpacs,floatfix,superscriptaddress]{revtex4}
\usepackage{graphicx}

\begin{document}
\flushbottom

\title{Temperature dependent tunneling spectroscopy in the heavy fermion CeRu$_{2}$Si$_{2}$ and in the antiferromagnet CeRh$_{2}$Si$_{2}$}
\author{A. Maldonado}
\affiliation{Laboratorio de Bajas Temperaturas, Departamento de
F\'isica de la Materia Condensada \\ Instituto de Ciencia de
Materiales Nicol\'as Cabrera, Facultad de Ciencias \\ Universidad
Aut\'onoma de Madrid, 28049 Madrid, Spain}
\author{H. Suderow}
\email[Corresponding author: ]{hermann.suderow@uam.es}
\affiliation{Laboratorio de Bajas Temperaturas, Departamento de
F\'isica de la Materia Condensada \\ Instituto de Ciencia de
Materiales Nicol\'as Cabrera, Facultad de Ciencias \\ Universidad
Aut\'onoma de Madrid, 28049 Madrid, Spain}
\author{S. Vieira}
\affiliation{Laboratorio de Bajas Temperaturas, Departamento de
F\'isica de la Materia Condensada \\ Instituto de Ciencia de
Materiales Nicol\'as Cabrera, Facultad de Ciencias \\ Universidad
Aut\'onoma de Madrid, 28049 Madrid, Spain}
\author{D. Aoki}
\affiliation{INAC, SPSMS, CEA Grenoble, 38054 Grenoble, France}
\author{J. Flouquet}
\affiliation{INAC, SPSMS, CEA Grenoble, 38054 Grenoble, France}

\begin{abstract}
CeRu$_{2}$Si$_{2}$ and CeRh$_{2}$Si$_{2}$ are two similar heavy fermion stoichiometric compounds located on the two sides of the magnetic quantum critical phase transition. CeRh$_{2}$Si$_{2}$ is an antiferromagnet below T$_{N}$=36 K with moderate electronic masses whereas CeRu$_{2}$Si$_{2}$ is a paramagnetic metal with particularly heavy electrons. Here we present tunneling spectroscopy measurements as a function of temperature (from 0.15 K to 45 K). The tunneling conductance at 0.15 K reveals V-shaped dips around the Fermi level in both compounds, which disappear in CeRu$_{2}$Si$_{2}$ above the coherence temperature, and above the N\'eel temperature in CeRh$_{2}$Si$_{2}$. In the latter case, two different kinds of V-shaped tunneling conductance dips are found.
\end{abstract}

 \date{\today} \maketitle

\section{INTRODUCTION}

Heavy fermion compounds show different ground states, e.g.
paramagnetic (PM) Kondo lattice or antiferromagnetic (AF) Fermi liquid, with
low critical temperatures that can be tuned by application of
pressure or magnetic
field\cite{Lohneysen07,Gegenwart08,FlouquetRoad}. Thus, they are key
references for zero temperature quantum phase transitions. In
particular, the so called magnetic quantum critical point (QCP)
results when switching at zero temperature from AF order to PM phase at the critical pressure
P$_{c}$. Parameters needed to describe this transition are the Kondo
temperature (T$_{K}$), the intersite magnetic correlations which
appear below T$_{corr}$, and the possible lift of the 4f
angular momentum degeneracy by the crystal field splitting
($\Delta_{CF}$). In addition, quasiparticles at the Fermi surface acquire its low temperature properties only when they are fully dressed below T$_{coh}$. During the last three decades, many studies sensing
macroscopic and microscopic aspects of these compounds have provided
new information about the behavior of a large number of them,
especially in 4f (Ce and Yb) and in 5f (U) intermetallic compounds\cite{Lohneysen07,Gegenwart08,FlouquetRoad}.
However, heavy fermion compounds are usually complex metals with
many bands crossing the Fermi surface, often giving "spaghetti" like
band structures (see \cite{Takashita96,Suzuki10} for
CeRu$_{2}$Si$_{2}$). Quantum oscillation experiments on the Fermi
surface demonstrate that heavy fermions coexist with light
itinerant carriers. The topology of the Fermi surface is generally well described by band structure
calculations\cite{Suzuki10}. However, the derived calculated
effective masses are often one or two orders of magnitude smaller
than those experimentally found in the areas of the Fermi surface
with heavy quasiparticles\cite{Takashita96}. It is now believed that
a large number of interesting effects appearing in these compounds
are intimately related to the Fermi surface with mixed heavy and
light electrons notably on Ce heavy fermion systems. Calculation methods have recently been improved and
are able to deal in more detail with band structure experiments, and
to help to understand atomic scale Scanning Tunneling Microscopy
(STM)\cite{Haule09}. The latter opens new perspectives, providing direct information about electronic band structure and correlations.
Successful STM measurements have been reported in PrOs$_4$Sb$_{12}$\cite{Suderow04}, PrFe$_4$P$_{12}$\cite{Suderow08Pr}, 
YbRh$_{2}$Si$_{2}$\cite{Ernst11}, URu$_{2}$Si$_{2}$\cite{Schmidt10,Aynajian10} and in CeRhIn$_{5}$ and
CeCoIn$_{5}$\cite{Aynajian12}.

Our aim is to realize scanning tunneling spectroscopy in the
two cerium tetragonal heavy fermion compounds CeRu$_{2}$Si$_{2}$ and
CeRh$_{2}$Si$_{2}$ which are respectively in paramagnetic and
antiferromagnetic ground state\cite{FlouquetRoad,Knafo10}. In Fig.
1(a) we schematically show the situation of both compounds in the
phase diagram of heavy fermions in terms of the tuning parameter
$\delta$, which describes the competition between the local Kondo effect
and the magnetic intersite interactions. In these compounds, the intersite
magnetic correlations prevail over the single site Kondo effect. Furthermore, the crystal field splitting is strong enough to deal with Cerium doublet levels. The properties of both
compounds are very well documented including de Haas Van Alphen measurements
(see \cite{Takashita96,Suzuki10} for CeRu$_{2}$Si$_{2}$ and
\cite{Araki01} for CeRh$_{2}$Si$_{2}$).

CeRu$_{2}$Si$_{2}$ has a paramagnetic ground state and is close to a
magnetic QCP. The electronic term in the specific heat, C/T,
strongly increases when reducing temperature\cite{FlouquetRoad,Besnus85,Fisher91}. The Fermi liquid
AT$^{2}$ law is found in the resistivity behavior only below
$T_A=$1 K\cite{Haen87}. Furthermore, the electronic Gr\"uneisen parameter
$\Omega_{e}$(T)\cite{Lacerda89} has also a strong temperature
variation at very low temperatures. At zero temperature, their
respective extrapolations give
$\gamma=(C/T)_{T\rightarrow0}\sim$350 mJmol$^{-1}$K$^{-2}$,
A$\sim$1 $\mu\Omega$cmK$^{-2}$, $\Omega_{e}$(T=0 K)$\sim$+200 and T$_{coh}\approx$ 9 K. The
QCP can be reached increasing the unit cell size by doping with La
or Ge, implying that it is located at a slightly negative pressure
of -0.3GPa\cite{Knafo09,Kambe97,Quezel88,Haen99}. Effective masses
range from 120m$_{0}$ down to the bare electron mass
m$_{0}$\cite{Takashita96}.

Although the molar volume of CeRh$_{2}$Si$_{2}$ is smaller than that
of CeRu$_{2}$Si$_{2}$\cite{Aoki12}, the ground state of
CeRh$_{2}$Si$_{2}$ is antiferromagnetic with a rather high N\'eel
temperature T$_{N}\sim$36 K\cite{Knafo10,Godart83}. The sublattice
magnetization is also rather large
(M$_{0}\sim$1.3$\mu_{B}$)\cite{Kawarazaki00}, generating a large
molecular field. C/T decreases on cooling with a relatively small
residual term
$\gamma\sim$23 mJmol$^{-1}$K$^{-2}$\cite{Graf97,Settai97}. The Fermi
liquid AT$^{2}$ term dominates the resistivity behavior already
below 10 K, with
A$\sim$1.4$\times$10$^{-3} \mu\Omega$cmK$^{-2}$\cite{Araki02,Boursier08}.
The electronic Gr\"uneisen parameter is negative with
$\Omega_{e}\lesssim$-20\cite{Graf97,Villaume08}. Specific heat and thermal expansion have sharp maxima at $T_N$. Fermi surface
experiments show effective masses ranging from 6m$_{0}$ to
0.36m$_{0}$\cite{Araki01,Suzuki10,SuzukiPrivate}. CeRh$_2$Si$_2$ is
a compensated metal with large carrier number, which is also the
case of CeRu$_2$Si$_2$ and YbRh$_{2}$Si$_{2}$\cite{Gegenwart08,Knebel08}, and with moderately
heavy and light electrons.

The effective mass of the heavy carriers scales with the magnitude
of $\gamma$ and roughly with the inverse of T$_{coh}$. In the PM side of the QCP, the temperature $T_A$ below
which the Fermi liquid AT$^{2}$ law is obeyed is far lower than $T_{coh}$. In
CeRh$_{2}$Si$_{2}$, antiferromagnetism with large T$_{N}$ is the
result of a strong interplay between local Kondo fluctuations and
magnetic intersite interactions, clearly enhanced by switching from
Ru to Rh ions. The effect of pressure is magnified by the proximity
to an intermediate valence regime associated with the inefficiency of crystal field splitting ($\Delta_{CF}$) when the Kondo temperature overpasses $\Delta_{CF}$. Antiferromagnetism disappears already
below 1 GPa\cite{Araki01,Araki02,Boursier08,Villaume08}.

The simple Doniach picture gives a Kondo temperature T$_{K}$=25 K
for CeRu$_{2}$Si$_{2}$ and T$_{K}$=50 K for
CeRh$_{2}$Si$_{2}$\cite{Severing89,Kawasaki98}. Specific heat and
susceptibility measurements in CeRu$_2$Si$_2$ give a doublet crystal
field ground state of Ising character located 200 K below the first
excited level\cite{Besnus85}. CeRh$_{2}$Si$_{2}$ also shows an
Ising doublet crystal field ground state. Neutron scattering and
susceptibility experiments suggest that the crystal field level is
between 200 K and 600 K above the ground
state\cite{Settai97,Severing89}. At first approximation, in the
paramagnetic phase of Ce heavy fermion compounds, it is assumed that the Kondo temperature governs
the high and intermediate temperature properties up to T$_{K}$ and
that the intersite interactions will play a role only below a
temperature T$_{corr}<$T$_{K}$. However, in CeRu$_{2}$Si$_{2}$
microscopic inelastic neutron scattering experiments\cite{Knafo09}
as well as macroscopic measurements\cite{Haen87} such as
magnetoresistivity, give T$_{corr}\sim$60 K; i. e., higher than
T$_{K}$. Thus the appearance of a resonance due to the interplay
between the initial localized 4f electrons and the light itinerant
(s, p, d) electrons is already renormalized by the intersite
interactions.

Tunneling into a system with localized states such as Kondo ions or
electrons with differing associated bands does not follow simple single
particle tunneling
theory\cite{Ternes09,Ernst11,Schmidt10,Aynajian10,Haule09,Hamidian11}.
The tunneling conductance is not proportional to the density
of states observed with macroscopic experiments, such as specific
heat. Instead, it is the result of interference effects between
the quasilocalized state, directly linked to the heavy carriers, and
the light electron band, which couples to the tip's light electron
states\cite{Ernst11,Schmidt10,Aynajian10}. A first approximation to
account for multiparticle tunneling effects is to consider coherent tunneling through two interfering channels (Fig. 1(b)). The
tunneling conductance can then be understood in terms of a Fano
lineshape\cite{Fano61,Li98,Madhavan98,Wahl04,Zhu12}. Depending on
the dominance of each channel, from preferential tunneling into the
quasilocalized states, to tunneling into the itinerant states,
different shapes with different asymmetry, ranging from a peak into
a dip, can be found in the tunneling conductance. As we show below, here
we mainly observe a symmetric dip located at the Fermi level. Symmetric dips have been observed in tunneling experiments on single Ce adatoms\cite{Li98}, and they are interpreted as preferential tunneling into the itinerant electron channel,
with a destructive interference to the quasilocalized ones,
which reduces the conductance at the resonant level. Therefore, the
tunneling conductance is given by an inverted Lorentzian function centered at
zero bias voltage
$g(V,T)=g_{off}+A\frac{(\frac{eV}{\Gamma})^{2}}{1+(\frac{eV}{\Gamma})^{2}}$,
where $\Gamma$ describes its width. This is equivalent to a modified Fano formula\cite{Fano61} with the asymmetry parameter
$q$=0 and the energy value where the resonance is centered
$\epsilon_{s}$=0. We take
$\Gamma$(T=0 K) to be of the order of the width of the heavy band, which inversely scales with the electronic effective mass
m$^{\ast}$ of the carriers in the heavy fermion compound.

\begin{figure}[ht]
\includegraphics[width=12cm,keepaspectratio, clip]{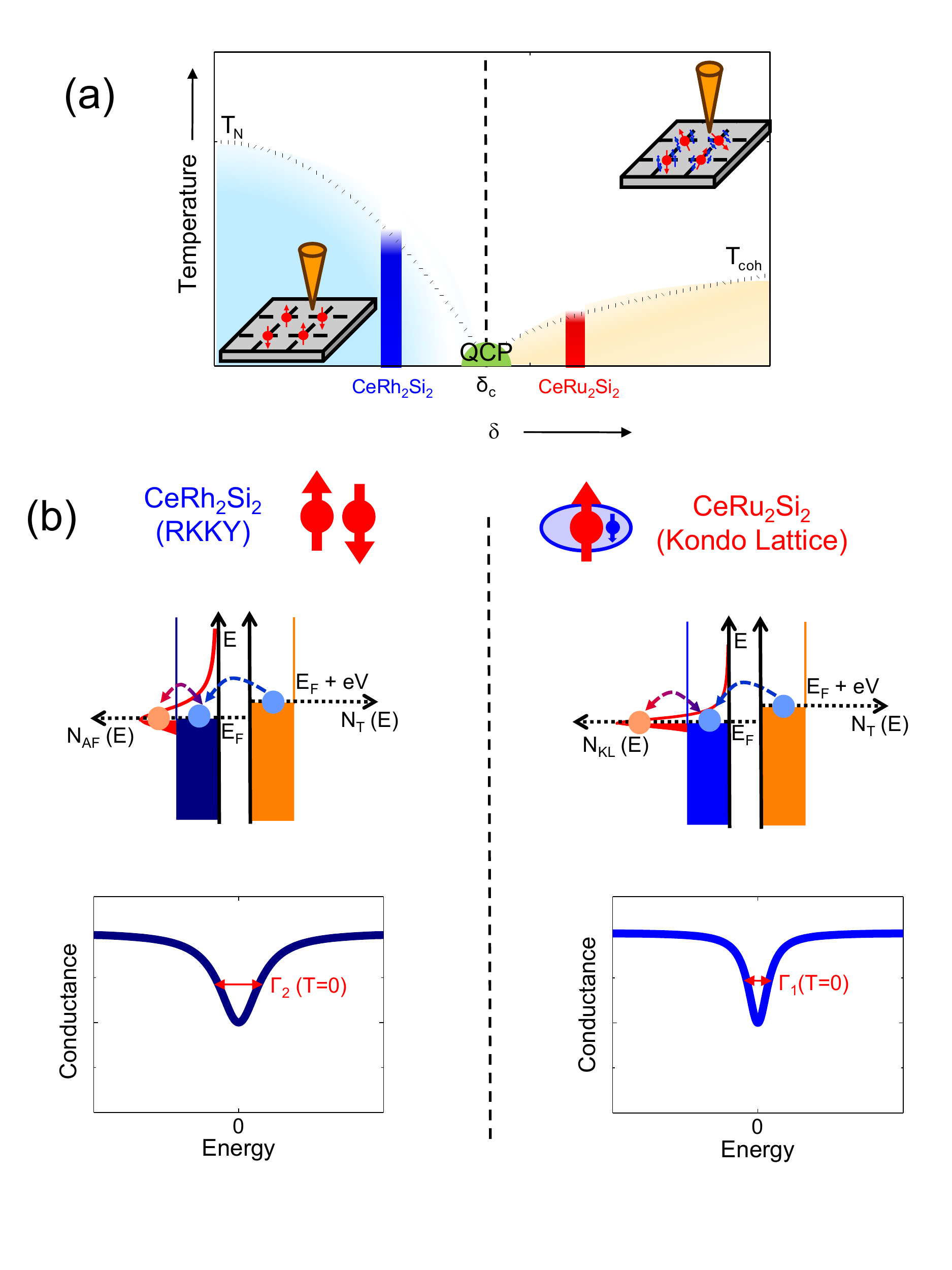}
\vskip -0cm \caption{(a) Sketch of the location of CeRu$_2$Si$_2$
and CeRh$_2$Si$_2$ within the T-$\delta$ phase diagram of heavy
fermions, where $\delta$ is the tuning parameter which describes the
competition between the local Kondo effect and the magnetic
intersite interactions. N\'eel (T$_N$) and coherence (T$_{coh}$)
temperatures are sketched as dashed black lines. The insets are cartoons of the experiment, with
an Au tip tunneling into a Kondo lattice (CeRu$_2$Si$_2$) and into
an antiferromagnet (CeRh$_2$Si$_2$). (b) Possible band diagrams for tunneling
processes between an Au tip into a system with localized Kondo
levels (left panel), and into a Kondo lattice (right panel) at zero
temperature. The predominant tunneling channel is represented by
blue dashed arrows, and the interaction with quasilocalized states by reddish
ones. The corresponding shape of the resulting tunneling conductance
curves at T=0 K is represented in the lower panels.} \label{Fig1}
\end{figure}

\section{EXPERIMENTAL DETAILS}

The experimental set-up consists of a home-made STM in a Oxford
Instruments MX400 dilution refrigerator with a positioning system
which allows to change the scanning window of 2x2 $\mu$m$^2$ in-situ
and without heating\cite{Suderow11}. We use tips of Au which are
prepared and cleaned in-situ as described in Ref. \cite{Rodrigo04b}.
We obtain the tunneling conductance $g(V,T)$ by numerically
derivating current vs bias voltage curves $I(V)$. $g(V,T)=dI/dV$ is normalized to the value obtained at
a bias voltage well above 20mV. Single crystal samples of
CeRu$_{2}$Si$_{2}$ and CeRh$_{2}$Si$_{2}$ were grown by Czochralski
method as in previous work (see e.g.\cite{Knafo10,Aoki12}). We broke
the samples along the basal plane of the tetragonal structure at
ambient conditions immediately before mounting them on the STM and
cooling down. Samples with a bright and optically flat surface were
selected. In general, we found surfaces with rather irregular shapes
(Fig.2) showing in some cases modulations at scales comparable to interatomic distances\cite{Raymond01,Grier84}. Of course, some amount of surface contamination is unavoidable. In some particular cases, this could significantly influence tunneling features. The features discussed here are however reproducible, and the observed temperature ranges where they appear coincide with temperature ranges known from macroscopic measurements. Moreover, we have changed the scanning window using the macroscopic positioning system, and present results obtained over clean surfaces showing reproducible imaging.

\begin{figure}[ht]
\includegraphics[width=14cm,keepaspectratio, clip]{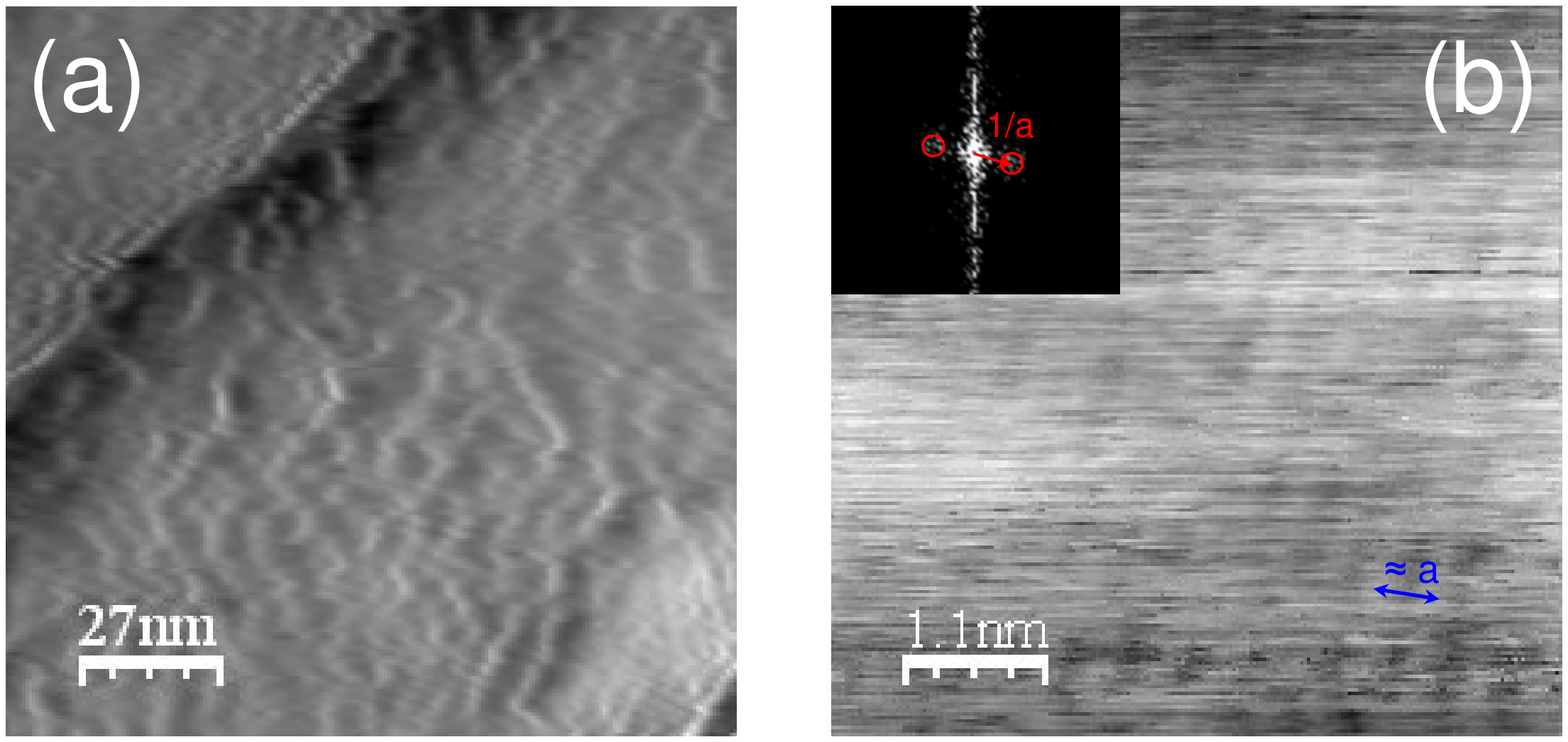}
\vskip -2cm \caption{Topography of CeRu$_{2}$Si$_{2}$ at 0.15 K taken at a conductance of 0.01 $\mu$S and a bias voltage of 50 mV in a scanning area of (a) 135 nm x 135 nm (grey scale corresponds to height changes by 3 nm) and (b) 5 nm x 5 nm (grey scale corresponds to height changes by 1 nm). Similar images are obtained in both compounds (CeRu$_2$Si$_2$ and CeRh$_2$Si$_2$). Inset shows the corresponding Fourier transform of the topography image, where the brightest Bragg peaks are highlighted by red circles. Their position in the reciprocal space, together with the period of the atomic scale features observed in the image, give a lattice parameter of about 0.4 nm.} \label{Fig2}
\end{figure}

\section{RESULTS AND DISCUSSION}

The tunneling conductance of CeRu$_{2}$Si$_{2}$ and
CeRh$_{2}$Si$_{2}$ at 0.15 K reveals features consisting of a sharp
V-shaped dip around zero bias voltage (Fig. 3). We find a different
behavior in both materials. The V-shaped dip is wider for
CeRh$_{2}$Si$_{2}$ than for CeRu$_{2}$Si$_{2}$. In CeRh$_{2}$Si$_{2}$ we find two differing characteristic behaviors, with deep and shallow minima
at zero bias, showing both roughly the same width. The dip
disappears at 9.5 K for CeRu$_{2}$Si$_{2}$, in good agreement with T$_{coh}$ measured with thermal expansion. It remains up to a
higher temperature, 45 K, for CeRh$_{2}$Si$_{2}$ (Fig.3). Fitting the conductance curves to the expression discussed above
$g(V,T)=g_{off}+A\frac{(\frac{eV}{\Gamma})^{2}}{1+(\frac{eV}{\Gamma})^{2}}$ we obtain the parameters discussed in Figs. 4 and 5. $\Gamma$ is the resonance width.
g$_{off}$ and A are the zero bias voltage conductance and the
amplitude of the dip, respectively. Both depend on the
tip-sample wavefunction coupling and on the density of states of the
sample at a given position. In Fig. 5 we show tunneling conductance curves for both compounds taken at different points of each sample surface, as well as the temperature dependence of the dip size $g_{off}$ and width $\Gamma$.

\begin{figure}[ht]
\includegraphics[width=12cm,keepaspectratio, clip]{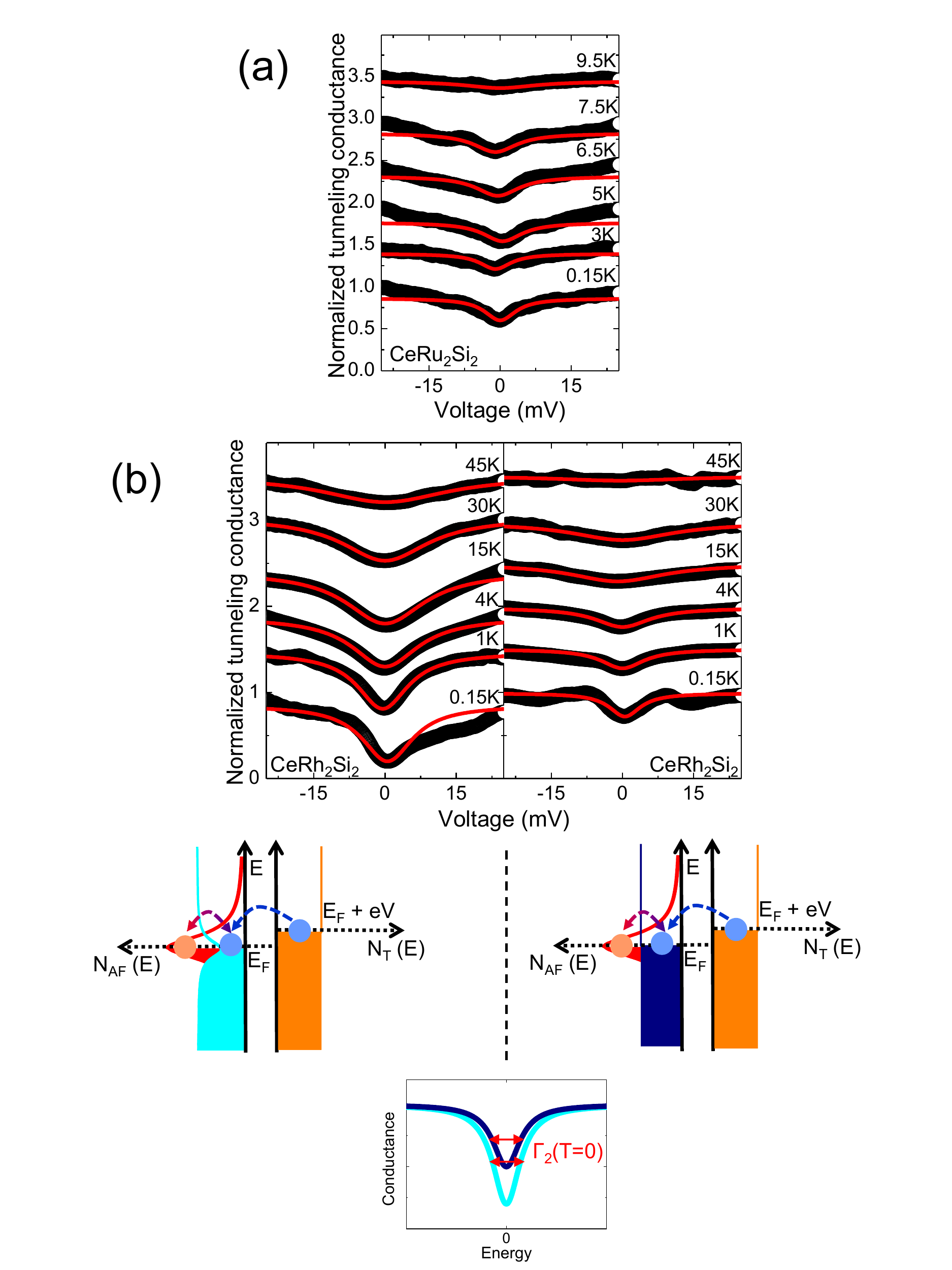}
\vskip -0cm \caption{The temperature evolution of the tunneling
spectroscopy curves for (a) CeRu$_{2}$Si$_{2}$ and (b)
CeRh$_{2}$Si$_{2}$ at an arbitrary position of their surfaces.
Curves have been shifted by 0.5 for clarity.  In the tunneling
conductance curves of CeRh$_{2}$Si$_{2}$ at 0.15 K we find two
characteristic behaviors, one with a deep V-shaped dip around zero
bias voltage (left panel) and another with a shallow one (right
panel). Red lines are fits to an inverted Lorentzian function as discussed in the
text. The corresponding parameters are given in Fig. 4. In the lower
panels of (b) band diagrams for tunneling processes between an Au
tip and CeRh$_{2}$Si$_{2}$ at zero temperature are shown similarly
as in Fig 1(b).} \label{Fig3}
\end{figure}

\begin{figure}[ht]
\includegraphics[width=12cm,keepaspectratio, clip]{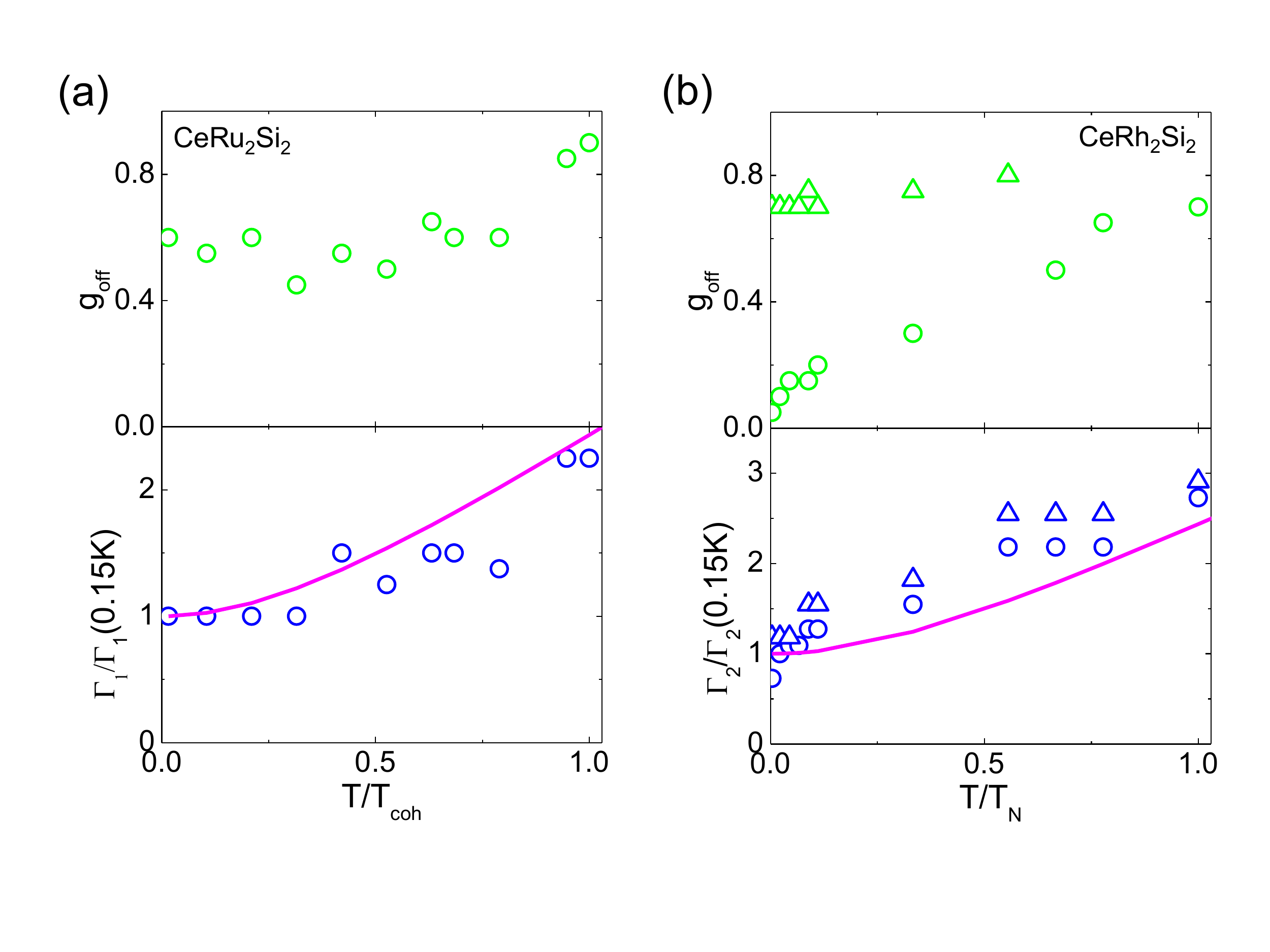}
\vskip -0cm \caption{Temperature evolution of the zero bias
voltage conductance g$_{off}$ and the width ($\Gamma$) of the dips
revealed in the tunneling spectra normalized to its lowest
temperature value for (a) CeRu$_{2}$Si$_{2}$ and (b)
CeRh$_{2}$Si$_{2}$ (triangles for the shallow dips and circles for the
strong dips observed in the tunneling conductance curves shown in
Fig. 3). Temperatures have been normalized to the corresponding
coherence (T$_{coh}\sim$9.5 K) and  N\'eel (T$_{N}\sim$45 K)
temperature for each compound. In the bottom panels we show (lines)
the Fermi liquid fit, using the normalized equation indicated in the
text.} \label{Fig4}
\end{figure}

\begin{figure}[ht]
\includegraphics[width=12cm,keepaspectratio, clip]{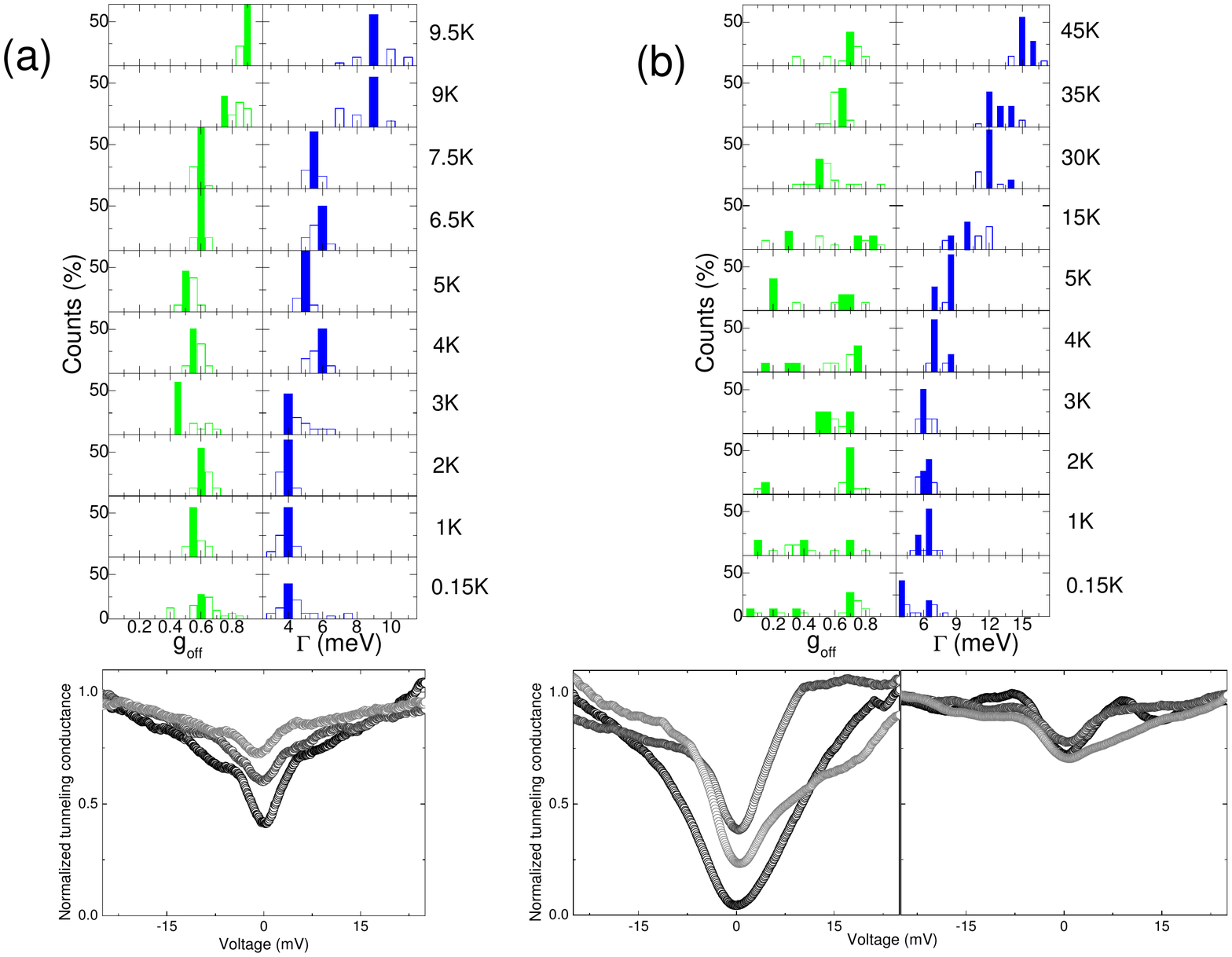}
\vskip -0cm \caption{Histograms for g$_{off}$ and $\Gamma$ obtained after fitting tens of tunneling conductance curves to the inverted Lorentzian function discussed in the text, which were taken at several different temperatures and at different positions of the surface of (a) CeRu$_{2}$Si$_{2}$ and (b) CeRh$_{2}$Si$_{2}$. Lower panels show examples of the different tunneling spectra found at 0.15K over the surface of (a) CeRu$_{2}$Si$_{2}$ and (b) CeRh$_{2}$Si$_{2}$.} \label{Fig5}
\end{figure}

The fits of the tunneling conductance curves at the lowest temperature (0.15 K) give a larger width $\Gamma$ for the dips observed in CeRh$_{2}$Si$_{2}$ ($\Gamma_{2}$(0.15 K) $\simeq$ 5.5 meV) than in CeRu$_{2}$Si$_{2}$ ($\Gamma_{1}$(0.15 K) $\simeq$ 4 meV). Moreover, in CeRh$_{2}$Si$_{2}$ the two
characteristic behaviors observed at 0.15 K, with a shallow and a
deep V-shaped dip, show roughly the same width $\Gamma_{2}$. The difference between both sets of curves is thus due to a different value of g$_{off}$.

In CeRu$_2$Si$_2$, g$_{off}$ changes with temperature only
weakly, but $\Gamma$ increases. In CeRh$_2$Si$_2$ we observe a
temperature variation of g$_{off}$ which is different depending on
the shape of the curves. For the shallow dips, g$_{off}$ do not vary
significantly with temperature, as in CeRu$_{2}$Si$_{2}$. However,
for the deep V-shaped dips, g$_{off}$ strongly increases. $\Gamma$
increases with temperature similarly in both kinds of curves.

In both systems, despite the different values observed at low
temperatures, $\Gamma$ increases similarly with temperature.
We can compare the thermal evolution of both $\Gamma_{1}$ and
$\Gamma_{2}$ with the Fermi liquid prediction for temperature
broadening of $\Gamma$ normalized to its zero temperature value,
$\Gamma$(T=0), which can be written as
$\Gamma/\Gamma(T=0)=\sqrt{\frac{1}{2}(\pi \frac{T}{T^{\ast}})^2+1}$\cite{Schiller00}. Here, we take T$^{\ast}$ as the temperature for which the V-shaped dip disappears in CeRu$_2$Si$_2$ (T$_{coh}$) and the
N\'eel (T$_N$) temperature for CeRh$_2$Si$_2$. We observe that the thermal broadening of $\Gamma$ for both compounds roughly follows the Fermi liquid prediction (bottom panel of Fig. 4). Therefore, the thermal smearing of the tunneling features is only determined by the characteristic energy scales for each compound, that are given by the corresponding values of $\Gamma_{1}$ and $\Gamma_{2}$ at 0.15K. Apart from the different energy scales obtained at the lowest temperatures, the temperature evolution of the tunneling features is roughly the same for CeRu$_{2}$Si$_{2}$ and CeRh$_{2}$Si$_{2}$.

The weak temperature dependence of g$_{off}$ observed in
CeRu$_2$Si$_2$ is similar to the one observed in the thermal evolution of the
tunneling spectra of URu$_{2}$Si$_{2}$ for temperatures above the
hidden order transition\cite{Aynajian10}. In CeRh$_2$Si$_2$ we clearly find two different behaviors for g$_{off}$ at different positions, which also evolve differently with temperature. This shows that long range magnetic order affects the tunneling signal. In CeRh$_{2}$Si$_{2}$ two antiferromagnetic sublattices appear at low temperature\cite{Kawarazaki00}. Possibly, additional gap opening or other features can give different tunneling conductance curves on specific surfaces. This can be re-inforced by different behavior in the magnetic correlation lengths, as obtained in neutron scattering experiments. In CeRh$_2$Si$_2$, it will rapidly reach atomic distances on cooling (after its divergence at T$_N$)\cite{Flouquet12}. In CeRu$_{2}$Si$_{2}$, which is closer to the QCP, the magnetic correlation length increases smoothly on cooling being a few atomic distances at very low temperatures\cite{Regnault90}. So that the Rh compound should be prone to show more local size surface dependent effects. It will be interesting to check the variations in the tunneling behavior when doping CeRu$_{2}$Si$_{2}$ with Rh\cite{Aoki12}, because it can unveil electronic features of magnetic interactions close to the quantum critical point.

\section{CONCLUSIONS}

In conclusion, we have measured the features in the tunneling
conductance curves of two Ce-based heavy fermion compounds
(CeRu$_{2}$Si$_{2}$ and CeRh$_{2}$Si$_{2}$) as a function of
temperature. We find V-shaped dips which signal heavy band formation in CeRu$_{2}$Si$_{2}$ and an antiferromagnetically ordered
phase in CeRh$_{2}$Si$_{2}$. The different
temperature evolution of the observed zero bias V-shaped dip
reflects the formation of different magnetic heavy-fermion ground
states.

The Laboratorio de Bajas Temperaturas is associated to the ICMM of
the CSIC. This work was supported by the Spanish MICINN and MEC
(Consolider Ingenio Molecular Nanoscience CSD2007-00010 program,
FIS2011-23488 and FPU grant), by the Comunidad de
Madrid through program Nanobiomagnet and by ERC (NewHeavyFermion),
and French ANR projects (CORMAT, SINUS, DELICE).


\end{document}